\newcommand{\MainColor}[1]{{#1}}

\newcommand{\SecQF}
	{\MainColor{\ref{app:QF}}}
\newcommand{\SecBiphotonTheory}
	{\MainColor{\ref{app:BiphotonTheory}}}
\newcommand{\SecHs}
	{\MainColor{\ref{app:Hs}}}
\newcommand{\SecHp}
	{\MainColor{\ref{app:Hp}}}
\newcommand{\SecEquivalence}
	{\MainColor{\ref{app:Equivalence}}}

\newcommand{\TableIII}
	{\MainColor{Table~\ref{table:three}}}

\newcommand{\EqBiphoton}
	{\MainColor{equation~(\ref{eq:biphoton})}}
\newcommand{\EqBiphotonEqFWM}
	{\MainColor{equations~(\ref{eq:biphoton})~and (\ref{eq:FWM})}}
\newcommand{\EqFWMtoEqEITm}
	{\MainColor{equations~(\ref{eq:FWM})-(\ref{eq:EITm})}}
\newcommand{\EqHs}
	{\MainColor{equation~(\ref{eq:hs})}}

\newcommand{\FigsSixAandB}
	{\MainColor{figures~\ref{fig:six}(a)~and \ref{fig:six}(b)}}

\newcommand{\AppColor}[1]{{#1}}

\newcommand{\FigsTwoAandB}
	{\AppColor{figures~\ref{fig:two}(a) and \ref{fig:two}(b)}}
\newcommand{\FigThreeA}
	{\AppColor{figure~\ref{fig:three}(a)}}
\newcommand{\FigThreeB}
	{\AppColor{figure~\ref{fig:three}(b)}}
\newcommand{\FigsThreeAandB}
	{\AppColor{figures~\ref{fig:three}(a) and \ref{fig:three}(b)}}
\newcommand{\FigsTwoThreeAll}
	{\AppColor{figures~\ref{fig:two}(a), \ref{fig:two}(b), \ref{fig:three}(a), and 
		\ref{fig:three}(b)}}
\newcommand{\FigFourAll}
	{\AppColor{figure~\ref{fig:four}}}
\newcommand{\FigFiveB}
	{\AppColor{figure~\ref{fig:five}(b)}}

\documentclass[12pt]{iopart}
\usepackage{cite}
\expandafter\let\csname equation*\endcsname\relax
\expandafter\let\csname endequation*\endcsname\relax
\usepackage[colorlinks=true, linkcolor=blue, citecolor=blue, urlcolor=blue]{hyperref}

\newcommand{\secRefs}{\section*{References}}
\newcommand{\ShowEmail}[1]{#1}
\newcommand{\onlinecite}[1]{\cite{#1}}

\newcommand{\FIGs}[1]{Figures~\ref{fig:#1}}

\newcommand{\RF}[1]{Reference~\cite{#1}}
\newcommand{\RFs}[1]{References~\cite{#1}}
\newcommand{\fig}[1]{figure~\ref{fig:#1}}
\newcommand{\figs}[1]{figures~\ref{fig:#1}}
\newcommand{\eq}[1]{equation~(\ref{eq:#1})}
\newcommand{\eqs}[1]{equations~(\ref{eq:#1})}
\newcommand{\rf}[1]{reference~\cite{#1}}
\newcommand{\rfs}[1]{references~\cite{#1}}

\usepackage{graphicx}
\usepackage{array}
\usepackage{amsmath, amssymb}
\usepackage{color}
\usepackage{ulem}

\newcommand{\CiteTableColdAtom}{
	HP2,
	HP7, 
	HP19, 
	SFWM.dLambda.ColdAtoms1.Du, 
	SFWM.dLambda.ColdAtoms2.Zhang, 
	OurAPL2022, 
	Table15}
\newcommand{\CiteTableHotAtom}{
	OurPRR2022, 
	Table9}

\newcommand{\CiteTableMicroResonatorWaveguideSPDC}{
	Table5, 
	Table8,
	Table16}
\newcommand{\CiteTableMicroResonatorWaveguideSFWM}{
	Table3, 
	Table10, 
	Table12, 
	Table13, 
	Table14, 
	Table17}
\newcommand{\CiteTableNonlinearCrystal}{
	Table2, 
	Table6, 
	SPDC.2017a, 
	SPDC.2016a, 
	Table19, 
	PeiYu22}
\newcommand{\CiteSPDC}{
	Table2,
	Table6,
	SPDC.2017a,
	SPDC.2016a,
	Table19,
	SPDC.2015b,
	PeiYu22,
	Table5, 
	Table8,
	Table16,
	SPDC.2017b,
	AddWaveguide2009,
	AddOpticalFiber2012,
	AddCrystal2012,
	AddCrystal2024}
\newcommand{\CiteSFWM}{
	HP2,
	HP7, 
	HP19, 
	SFWM.dLambda.ColdAtoms1.Du, 
	SFWM.dLambda.ColdAtoms2.Zhang, 
	OurAPL2022, 
	Table15,
	AddColdAtom2016,
	AddColdAtom2020,
	AddColdAtom2023,
	OurPRR2022, 
	Table9,
	Table30,
	Table27,
	OurQST2025,
	OurOpticaQ2025,
	AddHotAtom2020,
	AddHotAtom2022,
	Table3,
	Table10,
	Table12,
	Table13,
	Table14,
	Table17}
\newcommand{\CiteQualityFactor}{
	Table3, 
	Table10}

\begin{document}
\title{Fundamental limit on the heralded single photons' spectral brightness}

\author{
Tse-Yu Lin,$^1$
Wei-Kai Huang,$^1$ 
Pei-Yu Tu,$^1$
Yong-Fan Chen,$^2$
Ite A. Yu$^{1,3,\ShowEmail{*}}$}

\address{
$^1$Department of Physics and Center for Quantum Science and Technology, National Tsing Hua University, Hsinchu 30013, Taiwan \\
$^{2}$Department of Physics, National Cheng Kung University, Tainan 70101, Taiwan \\
$^3$National Center for Excellence in Quantum Information Science and Engineering, National Tsing Hua University, Hsinchu 30013, Taiwan
\ShowEmail{\\$^{*}$yu@phys.nthu.edu.tw}
}

\begin{abstract}
Heralded single photons (HSPs) are the versatile flying qubits in quantum communication and networks due to their ability to remove the randomness of arrival time and enhance the transmission reliability. As the generation rate of HSPs increases or their linewidth narrows, both of which are desirable for quantum information processing, the fundamental limit of spectral brightness (SB), defined as the generation rate per unit linewidth, remains unclear. To examine the existence and value of such a limit, we systematically studied the SB together with the cross-correlation function, or equivalently, the signal-to-background ratio (SBR). We ultimately derive an upper bound on SB that applies universally to all types of HSP sources. A newly defined quantity governs this limit, the quality factor, which is the product of SBR and effective SB. The quality factor indicates how closely an HSP source approaches an ideal noise-free source. Furthermore, by employing an HSP source based on hot atomic vapor, we achieved an SB of $(8.5\pm0.3)$$\times$$10^5$ pairs/s/MHz and a quality factor of $0.73\pm0.02$ under the single-photon criterion. Both values represent the highest reported performance to date among all HSP platforms. These results provide a unified benchmark for evaluating and optimizing HSP sources.
\end{abstract}

\newcommand{\FigOne}{
	\begin{figure}[t]
	\center{\includegraphics[width=78mm]{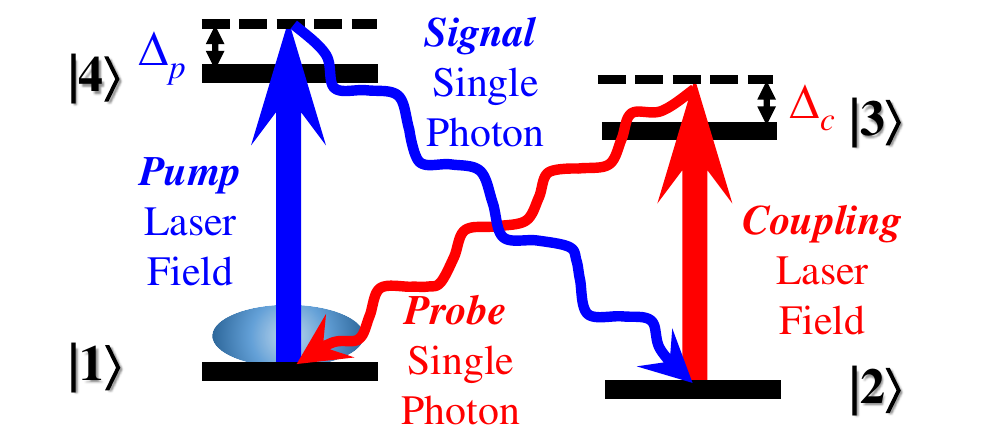}}
	\caption{
The double-$\Lambda$ transition scheme for the biphoton generation used in this work. We applied the pump and coupling fields to hot atomic vapor and produced the signal and probe photons. The pump field and signal photon (the coupling field and probe photon) form the first (the second) two-photon Raman transition with a one-photon detuning of $\Delta_p$ ($\Delta_c$).
	}
	\label{fig:transitions}
	\end{figure}
}
\newcommand{\FigTwo}{
	\begin{figure*}[t]
	\center{\includegraphics[width=156mm]{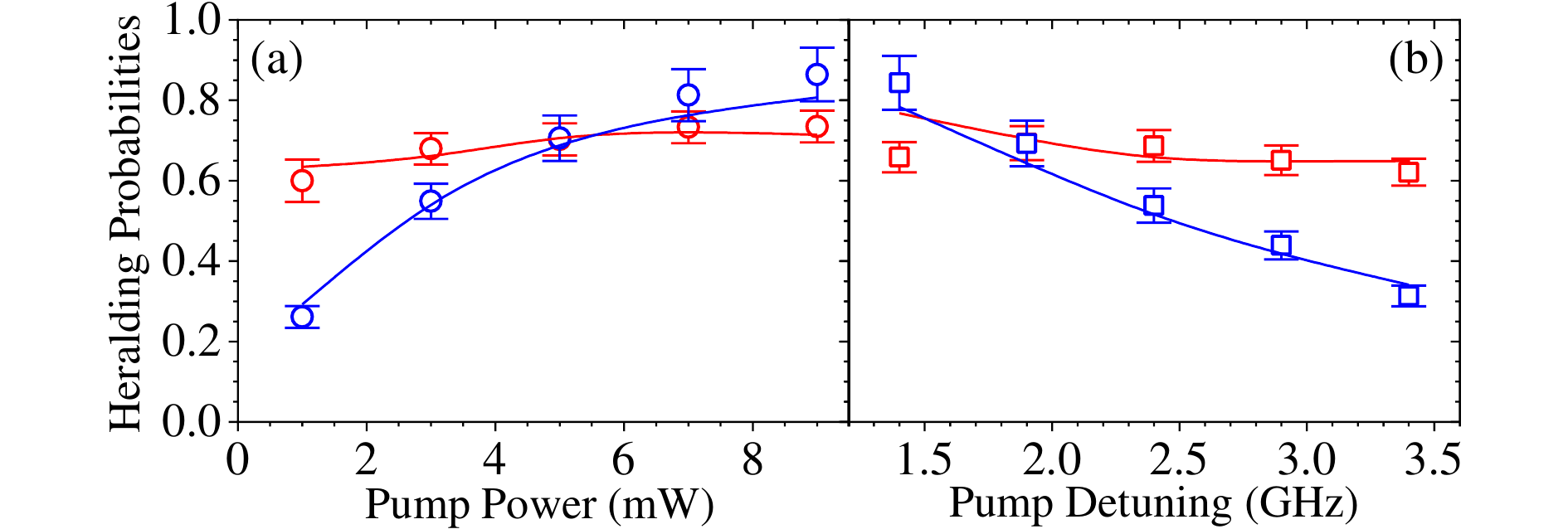}}
	\caption{
The signal's and probe's heralding probabilities $h_s$ and $h_p$ are shown as functions of the pump (a) power and (b) detuning. Red (and blue) circles or squares are the experimental data with the signal (and probe) photons to herald coincidence counts. Each line represents the theoretical prediction with a proportionality matching the count rate in theory to that in measurement. The coupling power and detuning were 17~mW and 1.00~GHz in all the measurements. The pump detuning was 2.90~GHz in (a), and the pump power was 2~mW in (b). In the theoretical calculation, we used $\alpha$ (OD) = 500, $\Omega_c =$ 12$\Gamma$, $\Omega_p$ = 2.8$\Gamma$$\times$$\sqrt{\rm pump~power~in~mW}$, $b =$ 0.375, and the various values of $\gamma$, which ranged 0.0072$\Gamma$$\sim$0.020$\Gamma$ in (a) and 0.0067$\Gamma$$\sim$0.019$\Gamma$ in (b). These calculation parameters can be found in \EqFWMtoEqEITm\ and their determination methods are described in \SecBiphotonTheory.
	}
	\label{fig:two}
	\end{figure*}
}
\newcommand{\FigThree}{
	\begin{figure*}[t]
	\center{\includegraphics[width=156mm]{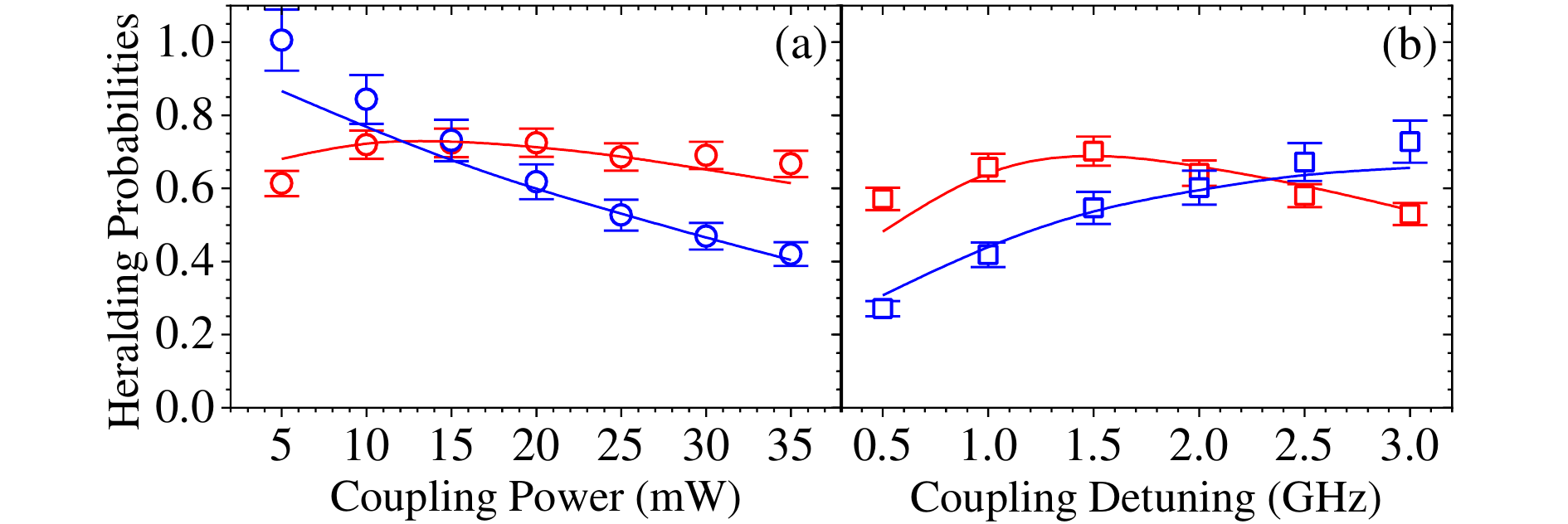}}
	\caption{
The signal's and probe's heralding probabilities $h_s$ and $h_p$ are shown as functions of the coupling (a) power and (b) detuning. Red (and blue) circles or squares are the experimental data with the signal (and probe) photons to herald coincidence counts. Each line represents the theoretical prediction with a proportionality matching the count rate in theory to that in measurement. The pump power and detuning were 2~mW and 1.90~GHz in all the measurements. The coupling detuning was 1.00~GHz in (a), and the coupling power was 35~mW in (b). In the theoretical calculation, we used $\alpha$ (OD) = 500, $\Omega_c =$ 3.0$\Gamma$$\times$$\sqrt{\rm coupling~power~in~mW}$, $\Omega_p$ = 4.0$\Gamma$, $b =$ 0.375 in (a) and 0.315 in (b), and the various values of $\gamma$, which ranged 0.0075$\Gamma$$\sim$0.045$\Gamma$ in (a) and 0.0089$\Gamma$$\sim$0.011$\Gamma$ in (b).
	}
	\label{fig:three}
	\end{figure*}
}
\newcommand{\FigFour}{
	\begin{figure}[t]
	\center{\includegraphics[width=156mm]{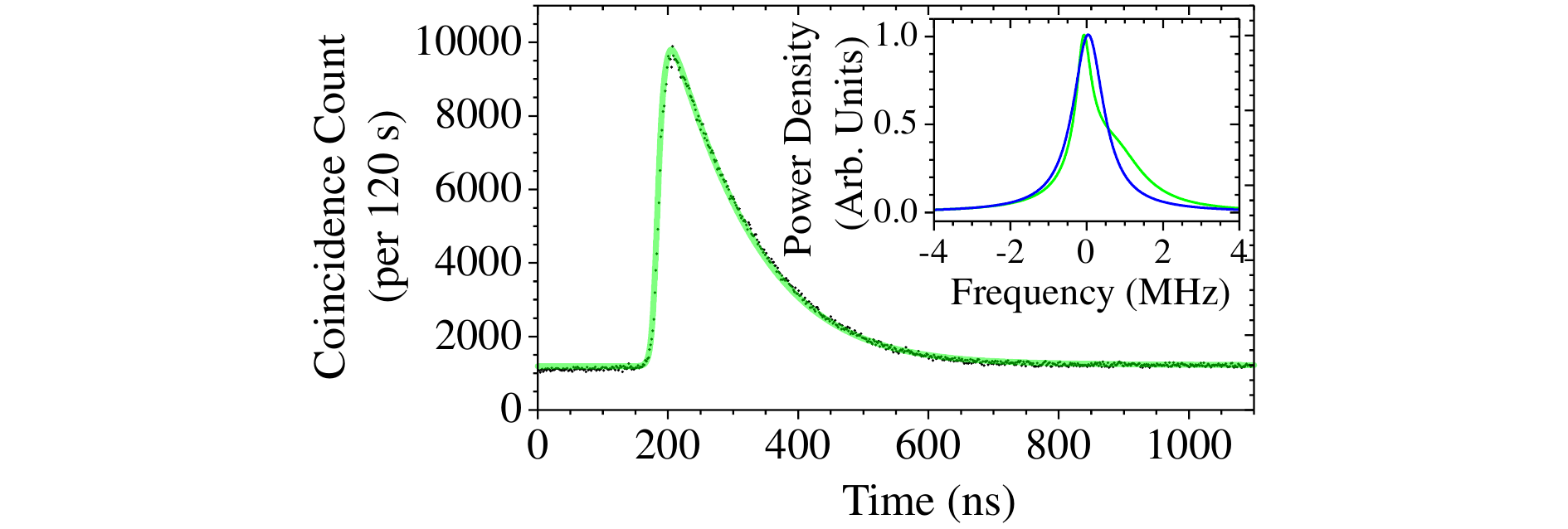}}
	\caption{
Biphoton wave packet or two-photon correlation function, i.e., coincidence count as a function of the delay time of the probe photon upon the signal photon's heralding. Black dots are the experimental data with the highest quality factor achieved in this work, and the green line is the theoretical prediction. The pump and coupling powers (detunings) were 5.0 and 13~mW (1.90 and 0.90~GHz), respectively. The OD of the hot atomic vapor was about 500. The biphotons, collected inside the single-mode PMFs immediately after the atomic vapor cell, had a generation rate of 8.4$\times$$10^5$ pairs/s, a temporal FWHM of 124~ns, and an SBR of 7.1. The measurements also show that the signal's and probe's heralding probabilities were 0.92 and 0.99. In the inset, the green line represents the theoretical prediction's frequency spectrum with a FWHM of 0.86~MHz, and the blue line is the best fit of a Lorentzian function with a FWHM of 0.99~MHz. We used $\alpha =$ 500, $\Omega_c =$ 10.5$\Gamma$, $\gamma =$ 0.012$\Gamma$, and $b$ = 0.375 in the theoretical calculation.
	}
	\label{fig:four}
	\end{figure}
}
\newcommand{\FigFive}{
	\begin{figure*}[t]
	\center{\includegraphics[width=156mm]{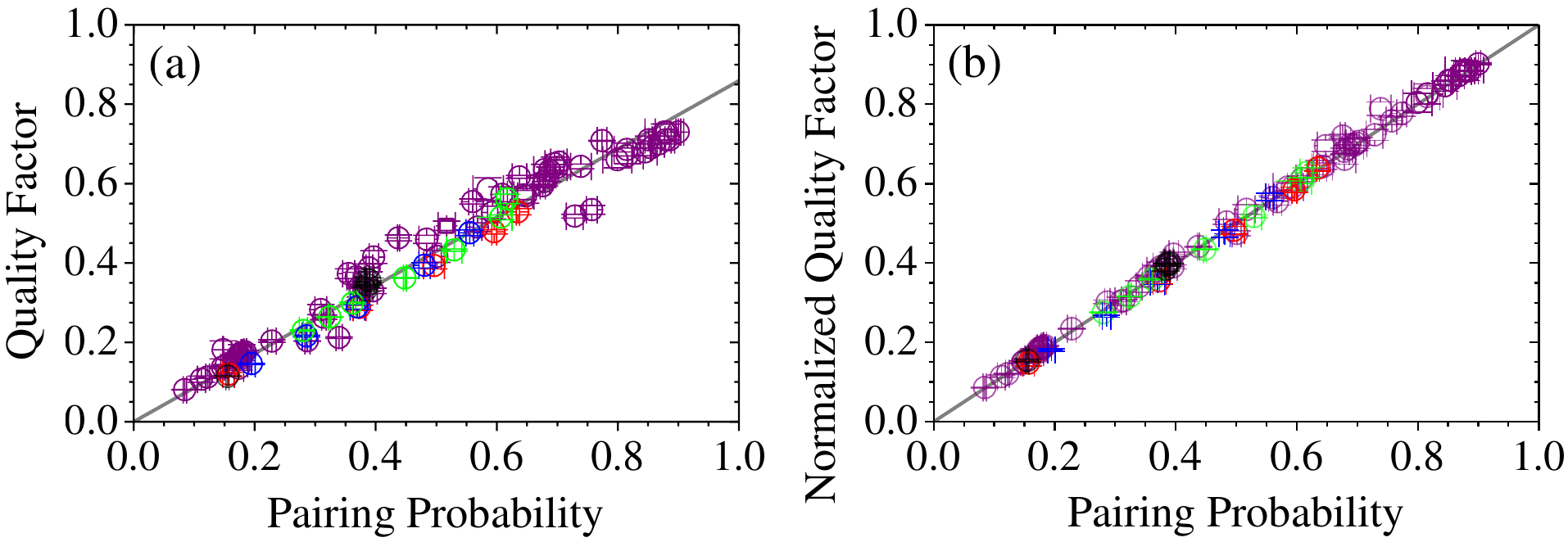}}
	\caption{
(a) The quality factor, $Q$, and (b) the normalized quality factor, $Q/C$, as functions of the pairing probability, $P$. The definitions of $Q$, $C$, and $P$ can be found in \eqs{relation}-(\ref{eq:definitionC}). Red, blue, green, and black circles are the experimental data relating to \figs{two}(a), \ref{fig:two}(b), \ref{fig:three}(a), and \ref{fig:three}(b), and purple ones are those during the search for the result in \fig{four}, where in the measurements the pump power ranged between 1 and 11~mW, the pump detuning changed from 1.4 to 3.4~GHz, the coupling power ranged between 3 and 36~mW, and the coupling detuning changed from 0 to 3.0~GHz. The straight line in (a) [or (b)] is the best fit (or theoretical prediction) with a slope of 0.92 (or 1).
	}
	\label{fig:five}
	\end{figure*}
}
\newcommand{\FigSix}{
	\begin{figure*}[t]
	\center{\includegraphics[width=156mm]{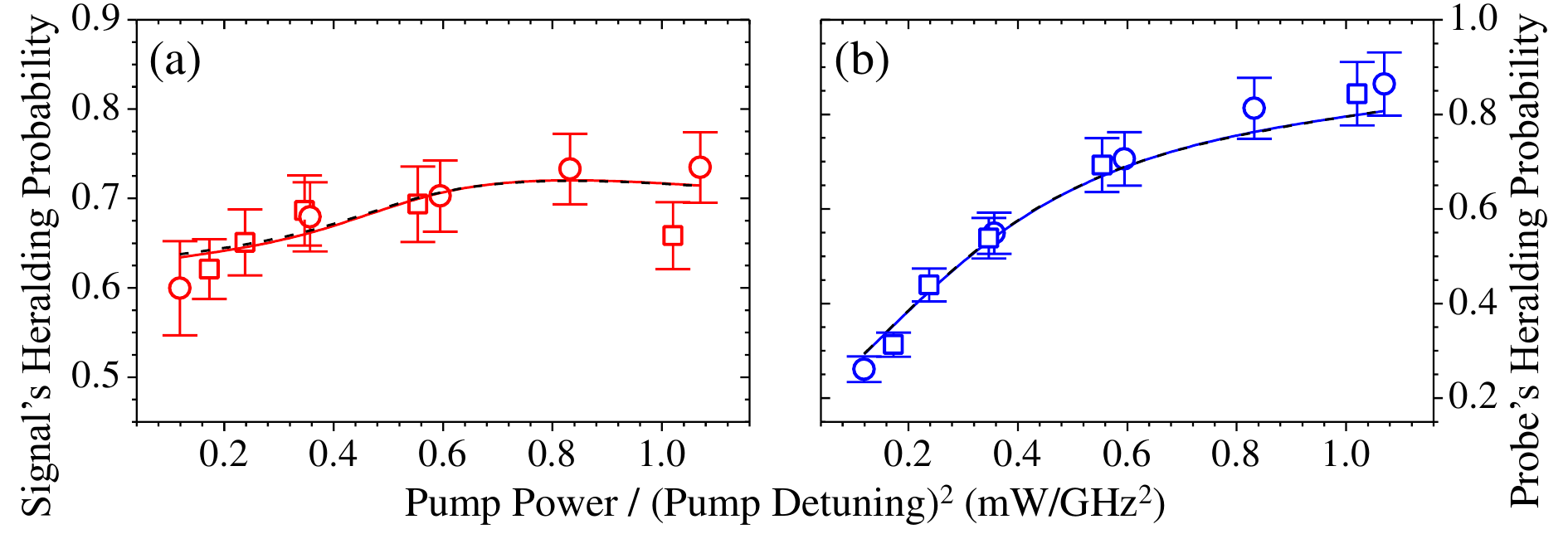}}
	\caption{
(a) The signal's and (b) the probe's heralding probabilities $h_s$ and $h_p$ are plotted as functions of the pump power divided by the square of pump detuning. Circles are the same experimental data in \fig{two}(a), and squares are the same experimental data in \fig{two}(b). Solid red and blue lines have the same theoretical values of $h_s$ and $h_p$ in \fig{two}(a), respectively. We calculated the dashed black lines by factoring the pump power and detuning out of the integral in \EqBiphoton.
	}
	\label{fig:six}
	\end{figure*}
}
\newcommand{\TableOne}{
	\begin{table*}[t]
	\caption{Biphoton sources with quality factors larger than 0.01.}
	\renewcommand{\arraystretch}{1.2}
	\renewcommand{\baselinestretch}{1} 
	\footnotesize
	\begin{tabular}{
		>{\raggedright\arraybackslash}p{18mm} 
		>{\raggedright\arraybackslash}p{26mm} 
		>{\raggedright\arraybackslash}p{28mm} 
		>{\raggedright\arraybackslash}p{28mm} 
		>{\raggedright\arraybackslash}p{16mm} 
		>{\raggedright\arraybackslash}p{14mm}
	}
	\hline
	\noalign{\vspace*{0mm}}
	Reference &
	Generation \newline Process &
	Platform &
	ESB \newline (pairs/s)$\cdot$$\mu$s &
	SBR &
	Quality \newline Factor \\
	\noalign{\vspace*{0mm}}
 	\hline 
	This work & SFWM & hot atoms & 
		$(1.03\pm0.03)$$\times$$10^5$$^\ast$ &
		7.1$\pm$0.1 & 0.73$\pm$0.02 \\
	\cite{Table2}&SPDC&nonlinear crystal&
		54$^\ast$&
		7.0$\times$$10^3$&0.38\\
	\cite{Table3}&SFWM&micro-resonator&
		170&
		2.0$\times$$10^3$&
		0.34\\
	\cite{PeiYu22}&SPDC&nonlinear crystal&
		1.8$\times$$10^4$&
		13&
		0.23\\
	\cite{Table5}&SPDC&waveguide&310&670&0.21\\
	\cite{Table6}&SPDC&nonlinear crystal&25&7.9$\times$$10^3$&0.20\\
	\cite{Table16}&SPDC&micro-resonator&
		630&
		260&
		0.16\\
	\cite{Table9}&SFWM&hot atoms&1.8$\times$$10^3$&42&0.076\\
	\cite{Table10}&SFWM&micro-resonator&720&100&0.072\\
	\cite{SPDC.2017a}&SPDC&nonlinear crystal&1.6$\times$$10^3$&41&0.066\\
	\cite{Table14}&SFWM&micro-resonator&
		300&
		220&
		0.066\\
	\cite{OurPRR2022}&SFWM&hot atoms&
		5.1$\times$$10^3$&
		12&
		0.061\\
	\cite{Table8}&SPDC&waveguide&1.4&4.0$\times$$10^4$&0.056\\
	\cite{Table12}&SFWM&micro-resonator&40&760&0.030\\
	\cite{Table19}&SPDC&nonlinear crystal&
		27&
		900&
		0.024\\
	\cite{Table17}&SFWM&micro-resonator&52&420&0.022\\
	\cite{SPDC.2016a}&SPDC&nonlinear crystal&42&290&0.012\\
	\noalign{\vspace*{1mm}}
	\hline
	\noalign{\vspace*{1mm}}
	\cite{Table15}&pulsed SFWM&cold atoms&
		38 (1.9$^\dagger$)&1.5$\times$$10^4$&0.57\\
	\cite{HP2}&pulsed SFWM&cold atoms&
		4.3$\times$$10^4$ (170$^\dagger$)&
		13&
		0.56\\ 
	\cite{HP19}&pulsed SFWM&cold atoms&
		7.8$\times$$10^3$ (780$^\dagger$)&22&0.17\\ 
	\cite{SFWM.dLambda.ColdAtoms2.Zhang}&pulsed SFWM&cold atoms&
		2.6$\times$$10^3$ (260$^\dagger$)&44&0.11\\ 
	\cite{OurAPL2022}&pulsed SFWM&cold atoms&
		1.9$\times$$10^3$(15$^\dagger$)&40&0.076\\ 
	\cite{HP7}&pulsed SFWM&cold atoms&
		2.8$\times$$10^3$(140$^\dagger$)&20&0.056\\ 
	\cite{SFWM.dLambda.ColdAtoms1.Du}&pulsed SFWM&cold atoms&
		2.2$\times$$10^3$ (220$^\dagger$)&24&0.053\\ 
	\hline
	\end{tabular}
	\hspace*{-7mm}
	\parbox[t]{162mm}{
		\begin{description}
			\setlength{\parskip}{0mm}
			\setlength{\itemsep}{0mm}
			\setlength{\labelsep}{0mm}
			\setlength{\itemindent}{-2mm}
			\item[$^\ast$] 
				When deriving generation rates from biphoton detection rates, 
				coupling efficiencies of the optical fibers used to collect single 
				photons are not included in the overall detection efficiencies. See 
				the second paragraph from the bottom of Sec.~\ref{sec:setup}.
			\item[$^\dagger$] 
				The average value of ESB, calculated with the consideration of 
				duty cycle, is shown in each parenthesis.
		\end{description}
	}
	\label{table:one}
	\end{table*}
}
\newcommand{\TableThree}{
	\begin{table*}[t]
	\caption{The quality factor's limit, i.e., $Q$, at the pairing probability $P$ 
		equal to 1.}
	\renewcommand{\baselinestretch}{1} 
	\renewcommand{\arraystretch}{1.8} 
	\footnotesize
	\begin{tabular}{
		>{\raggedright\arraybackslash}p{32mm} 
		>{\raggedright\arraybackslash}p{75mm} 
		>{\raggedright\arraybackslash}p{28mm}
        }
	\hline
	\noalign{\vspace*{0mm}}
	Function Name &
	Functional Form &
	Limit of $Q$ \\
	\noalign{\vspace*{0mm}}
    	\hline
	\parbox[t]{32mm}{\raggedright Single Exponential} & 
	\parbox[t]{75mm}{\raggedright $(1/\tau_w)\exp[-(\tau-\tau_0)/\tau_w]~(\tau 
		\geq \tau_0)~~{\rm and}~~0~(\tau < \tau_0)$} &
	\parbox[t]{28mm}{\raggedright $\ln 2 \approx 0.69$}\\ 
	\parbox[t]{32mm}{\raggedright Double Exponential} &
	\parbox[t]{75mm}{\raggedright $(1/2\tau_w)\exp[-|\tau-\tau_0|/\tau_w]$} &
	\parbox[t]{28mm}{\raggedright $\ln 2 \approx 0.69$}\\
	\parbox[t]{32mm}{\raggedright Gaussian} & 
	\parbox[t]{75mm}{\raggedright $(1/\sqrt{\pi}\tau_w)
		\exp[-(\tau-\tau_0)^2/\tau_w^2]$} &
	\parbox[t]{28mm}{\raggedright $2\sqrt{\ln 2/\pi} \approx 0.94$}\\
	\parbox[t]{32mm}{\raggedright Square} & 
	\parbox[t]{75mm}{\raggedright $1/\tau_w~(\tau_0 \leq \tau \leq \tau_0
		+\tau_w)~~{\rm and}~~0~({\rm otherwise})$} &
	\parbox[t]{28mm}{\raggedright $1.0$}\\
	\parbox[t]{32mm}{\raggedright This work} & 
	\parbox[t]{75mm}{\raggedright The biphoton temporal profile in \FigFourAll} &
	\parbox[t]{28mm}{\raggedright $0.81$}\\
	\hline
	\end{tabular} 
	\label{table:three}
	\end{table*}
}
\section{Introduction}

Long-distance quantum communication enables the secure transmission of information and the distribution of quantum correlations over extended distances, forming the backbone of quantum networks and quantum internet architectures. Such capabilities are essential for applications including quantum key distribution and distributed quantum computing. The realization of these protocols critically relies on the availability of high-quality single-photon sources. However, producing single photons as carriers of quantum information on demand is not easy because of the stochastic nature of their generation. There are two distinct types of approaching single-photon sources: (quasi)deterministic sources employing single emitters and heralded, probabilistic sources employing ensembles of emitters. The latter sources generate time-correlated photon pairs, termed biphotons, at random times. One photon of a pair is used to herald the presence of another, and they are called the heralding and heralded photons. The biphoton or heralded single photon (HSP) ensures that the arrival time of a single photon or qubit is known, thereby removing the randomness of the generation process and enhancing both the success rate and the reliability of the transmitted signal. Biphotons or HSPs can be generated via the processes of spontaneous parametric downconversion (SPDC) \cite{\CiteSPDC} or spontaneous four-wave mixing (SFWM) \cite{\CiteSFWM} in diverse media such as nonlinear crystals, waveguides, optical fibers, microresonators, laser-cooled atom clouds, and heated atomic vapor.

Biphoton sources have three important figures of merit: pairing probability, spectral brightness (SB), and signal-to-background ratio (SBR). The pairing probability $P$, a less commonly measured figure of merit, quantifies the likelihood of correlated single photons being in the heralding and heralded channels. And it is defined as
\begin{equation}
\label{eq:definitionP}
	P \equiv h_s h_p = (R_b/R_s) (R_b/R_p),
\end{equation}
where $h_s$ and $h_p$ represent the probability of detecting the heralded photon given that its partner photon is used as the herald, $R_b$ is the biphoton generation rate, and $R_s$ (or $R_p$) is the photon rate in the signal (or probe) detection channel. Note that we refer to the two photons in a biphoton pair as the signal and probe photons. In addition, a larger $P$ value indicates a smaller fraction of noise photons, which can be single photons due to uncorrelated emissions, correlatively emitted single photons that lose their counterparts in a medium, photons due to laser leakage, etc. When the $P$ value of a biphoton source approaches 1, the percentages of noise photons in the heralding and heralded channels approach 0, i.e., the source is very close to an ideal noise-free source. 

The definition of SB is the biphoton generation rate divided by the linewidth, which is the full width at half maximum (FWHM) of the biphoton wave packet's spectral profile. A biphoton source with a higher generation rate can produce information carriers more quickly. Given a finite bandwidth of light-matter interaction, the efficiency of a quantum operation increases when biphotons of a narrower linewidth are employed. Thus, the SB strongly influences the success rate of quantum operations. On the other hand, the SBR is defined as the ratio of the biphoton wave packet's peak height to the background level or, equivalently, the maximum of the cross-correlation function subtracted by 1, i.e., ${\rm Max}\left[ g^{(2)}_{sp}(\tau) \right] -1$ where the subscript $sp$ means the signal and probe photons and $\tau$ is the delay time between the two photons. The SBR is similar to the coincidence-to-accidentals ratio (CAR) \cite{CAR}. It affects the fidelity of quantum operations that utilize these HSPs. However, SB and SBR are not independent. A higher SB (or SBR) consistently results in a lower SBR (or SB) \cite{OurPRR2022}, and in recent decades, progress has been made in optimizing them \cite{SPDCreview, IPCreview, OurAQT2024}.

To elucidate the relationships among these figures of merit and to determine what value is the fundamental limit on the SB, we define a new quantity, the quality factor $Q$ as the product of the effective spectral brightness (ESB) and the SBR (see the next section). This quantity characterizes the combined performance of a biphoton source in terms of the SB and SBR. We showed that the $Q$ value is proportional to the pairing probability, with the proportionality determined by the waveform shape. Even in the limit of negligible background noise, the $Q$ value accurately accounts for the reduced SBR due to the overlap of successive photons as the photon generation rate or temporal width increases. Because the quantities used to define the quality factor are commonly measured in experiments, this definition provides a convenient way to quickly assess how closely a biphoton source approaches the ideal noise-free condition. In other words, a fundamental limit on the SB does exist because the pairing probability must be less than one.

In this paper, we will present a systematic study on the quality factor $Q$ and pairing probability $P$ of a double-$\Lambda$ SFWM biphoton source with hot atomic vapor. First, the universal relationship between $Q$ and $P$, outlined in \eq{relation}, will be introduced. Then, we will report how various experimental parameters affect $P$ and provide theoretical interpretations of the observed behaviors. Next, we will show that after optimizing various experimental parameters, a $P$ value of $0.89\pm0.02$ is achieved; i.e., approximately $94\%$ of the photons in the two detection channels are, on average, the paired biphotons. This pairing probability results in a quality factor of $0.73\pm0.02$, corresponding to an ESB of $(1.03\pm0.03)\times10^5$ (pairs/s)$\cdot$$\mu$s or an SB of $(8.5\pm0.3)\times10^5$ pairs/s/MHz at an SBR of $7.1\pm0.1$, referring to the biphotons collected inside the single-mode optical fibers immediately after the atomic vapor cell. These findings provide new insights into the fundamental imit of SB and establish guidelines for optimizing their SB given an SBR.

\section{Quality Factor of a Biphoton Source} \label{sec:QF}

In the following section, we motivate our definition of the quality factor $Q$. The effective spectral brightness (ESB), defined as $\Delta\tau R_b$, is an equivalent representation of the SB, defined as $R_b/\Delta f$, where $\Delta\tau$ ($\Delta f$) is the temporal (spectral) FWHM of a biphoton wave packet. Since $1/\Delta\tau$ is proportional to $2\pi\Delta f$, the ESB and SB differ by a factor on the order of $2\pi$. We chose the ESB to represent the SB for two reasons. First, the ESB and SBR form a simple, physically transparent relationship with the pairing probability. Second, because the ESB is defined in terms of temporal width, it is more straightforward to measure experimentally and to compare with results reported in the literature. Consider an ideal biphoton source that produces only pairs of single photons, with no noise. The wave packets of biphotons from an ideal noise-free source can produce background photons when two detected single photons, one in the heralding channel and the other in the heralded channel, originate from different pairs. The probability of wave-packet overlap increases with higher ESB, resulting in a higher background photon count rate (i.e., a lower SBR). With an ideal noise-free biphoton source, $Q$ reaches an upper limit, as do the ESB and SB under a given SBR.

To understand the relationship between the ESB and SBR, we derived a universal formula that includes both parameters (see \SecQF). To provide clearer physical insight, we show the formula by relating $Q$ (i.e., the product of the ESB and SBR) to the pairing probability $P$ of having correlated single photons in heralding and heralded channels as follows:
\begin{eqnarray}
\label{eq:relation}
	Q~{\rm(or~ESB \times SBR)} = C \times P, \\
\label{eq:definitionC}
	C \equiv \frac{f(\tau_0) \Delta\tau}{\int^{\infty}_{-\infty} f(\tau) d\tau},
\end{eqnarray}
where $C$ is a constant depending only on the shape of the biphoton wave packet, and $f(\tau)$, $\Delta\tau$, and $\tau_0$ represent the functional form, temporal FWHM, and peak position of the wave packet, respectively.

The value of the $Q$ is between 0 and $C$, where $C \leq 1$. The upper limit of $Q$, i.e., $C$, differs under various functional forms (see \SecQF~and \TableIII). For an ideal noise-free HSP or biphoton source, $P = 1$, i.e., $Q$ reaches its upper limit. An HSP source with a $Q$ value less than $C$ indicates that $P$ is less than 1 or, equivalently, the source contains noise photons. A lower $P$ value represents a higher ratio of noise photons to biphotons. According to \eq{relation}, $Q$ is linearly proportional to $P$ and is more easily measured. Therefore, the $Q$ value is a good metric indicating how closely an HSP source approaches an ideal noise-free one, while also being much easier to measure experimentally.

In Table~\ref{table:one}, we list the references that report a quality factor $Q\geq 0.01$ under the single-photon criterion, i.e., the HSP's conditional auto-correlation $<$ 0.5 \cite{Criterion} or, equivalently, SBR $>$ 6.5 \cite{OurOPEX2024}. This cutoff is chosen because sources with smaller $Q$ values are nearly two orders of magnitude lower than that achieved in this work, and the quality factors of references are derived from the reported $R_b$, SBRs, and temporal widths. Note that we omit \rf{SPDC.2015b} from the table because its biphoton source operating in pulse mode had a Q value that, to our knowledge, exceeds the physical bounds of this metric. All the biphoton sources listed in the table comply with the single-photon criterion. These studies, which involve different generation processes across various media or platforms, reveal that higher ESB (SBR) generally results in lower SBR (ESB). An interesting trade-off is also observed: solid-state systems offer higher SBRs with shorter temporal widths, whereas atomic systems provide longer temporal widths but lower SBRs. A similar trade-off also occurs between the SBR and the generation rate, and both trade-offs are encoded in \eq{relation}.

\TableOne

In \rfs{\CiteTableColdAtom}, the authors employed laser-cooled atoms to generate biphotons via the SFWM process. Their results typically exhibited the biphoton wave packets with long temporal widths due to the low decoherence rates in the cold-atom systems. Thus, $\Delta\tau$ exceeding 500 ns was achieved in several of these references. A cold-atom source operates in the pulse mode with a low duty cycle; during each period, the biphoton generation occupies only a small portion of the time, while the cooling and trapping of atoms occupy the remaining time. Consequently, the average $R_b$ and ESB values of a cold-atom source are low. The peak ESBs in these references ranged from 38 to 1.6$\times$$10^4$ (pairs/s)$\cdot$$\mu$s, while the corresponding SBRs were between 1.5$\times$$10^4$ and $13$, respectively. In \rfs{\CiteTableHotAtom}, the authors employed hot atomic vapor to generate biphotons via the SFWM process. Compared with cold-atom sources, hot-atom sources typically have shorter temporal widths but greater generation rates. The ESB of a hot-atom source usually surpasses the average ESB of a cold-atom source, which typically has a low duty cycle. Moreover, the ESB in this work surpassed even the best value of the peak ESBs of the cold-atom biphoton sources \cite{\CiteTableColdAtom}. Across these prior works and the present work, the reported ESBs ranged from 1.8$\times$$10^3$ to 9.8$\times$$10^4$ (pairs/s)$\cdot$$\mu$s, while the corresponding SBRs were between 42 and 6.9, respectively.

In \rfs{\CiteTableMicroResonatorWaveguideSPDC} and \onlinecite{\CiteTableMicroResonatorWaveguideSFWM}, the authors fabricated a microresonator or a waveguide, which enhanced the nonlinear light-matter interactions, on a chip. They applied a pump laser field to the microresonator or waveguide and generated biphotons via SPDC or SFWM. These sources have the benefits of a compact form and a low pump laser power. The ESBs reported in these references ranged from 1.4 to 720 (pairs/s)$\cdot$$\mu$s, while the corresponding SBRs were between 4.0$\times$$10^4$ and 77, respectively. In \rfs{\CiteTableNonlinearCrystal}, the authors used a nonlinear crystal and a pump laser field to generate biphotons via the SPDC process. They typically installed an optical cavity to narrow the biphoton linewidth, effectively prolonging its temporal width. However, the optical cavity's resonance frequency is sensitive to temperature, and its linewidth is not tunable. In these references, the reported ESBs ranged from 19 to 9.8$\times$$10^3$ (pairs/s)$\cdot$$\mu$s, while the corresponding SBRs were between 7.9$\times$$10^3$ and 13, respectively.

Since we derived \eq{relation} based on fundamental theory without utilizing any device- or system-dependent assumption, it applies to all studies. Moreover, because the definition of $Q$ relies solely on the quantities $R_b$, $\Delta\tau$, and SBR, which are routinely measured in experiments, it enables straightforward, consistent comparisons across different biphoton platforms.

\section{Experimental Setup} \label{sec:setup}

We generated biphotons via the SFWM process with a paraffin-coated cylindrical glass vapor cell (Precision Glassblowing TG-ABRB-I87-P) filled with isotopically enriched $^{87}$Rb atoms. The cell has a length of 75~mm and a diameter of 25.4~mm. Since the optical depth (OD), denoted by $\alpha$ in \EqFWMtoEqEITm, can affect the generation rate and temporal width of biphotons, we stabilized the cell temperature around 48 $^\circ$C to maintain a constant OD of about 500. 

The energy levels and optical transitions relevant to the experiment are shown in \fig{transitions}, where states $\lvert1\rangle$, $\lvert2\rangle$, $\lvert3\rangle$, and $\lvert4\rangle$ represent $\lvert5S_{1/2}, F = 2\rangle$, $\lvert5S_{1/2}, F = 1\rangle$, $\lvert5P_{1/2}, F = 2\rangle$, and $\lvert5P_{3/2}, F = 2\rangle$, respectively. The frequency separation between ground states $\lvert1\rangle$ and $\lvert2\rangle$ is approximately 6.8~GHz, while excited states $\lvert3\rangle$ and $\lvert4\rangle$ have the spontaneous decay rate $\Gamma$ of about 2$\pi$$\times$6 MHz. To prepare the population in the $\lvert1\rangle$ ground state, we applied a hyperfine optical pumping (HOP) field of roughly 10~mW in the form of a donut-shaped beam, which replaced the population in $\lvert2\rangle$ to $\lvert1\rangle$ while minimizing disturbance to the SFWM interaction region. 

\FigOne

The pump and coupling beams were generated by two external-cavity diode lasers (Toptica DL DLC pro) operating at 780 and 795 nm, respectively. We aligned the beams so that they propagate in the same direction as the collection direction of the signal and probe photons. The two laser beams completely overlapped within the vapor cell. This all-copropagating configuration ensured phase matching, thereby increasing four-wave-mixing efficiency; it also markedly reduced the ground-state decoherence rate caused by Doppler broadening. The pump and coupling fields were linearly polarized with the $p$ and $s$ polarizations, respectively. 

We detected the signal and probe photons in the $s$- and $p$-polarization directions, using two Excelitas SPCM-AQRH-13-FC single-photon counting modules (SPCMs) with dark count rates of approximately 200 counts/s. To eliminate leakage of the pump and coupling laser fields into the SPCMs, we used polarization filters (Thorlabs GTH10M-B) and etalon filters (Quantaser FPE002). These filters suppressed the laser fields by more than 130 dB; thus, the pump and coupling leakages were approximately 19 and 5 counts/s/mW, respectively. The leakage contributions in both channels account for less than 1\% of the total counts and are therefore negligible. 

The signal and probe photons were collected by two single-mode polarization-maintaining optical fibers (PMFs), Thorlabs P3-780PM-FC-2, which also collected 75\% and 69\% of the pump and coupling beams, respectively. The overall detection efficiencies, including the SPCM quantum efficiency, etalon and polarization filter transmittances, and other optical losses but excluding the PMFs' collection efficiencies, were $D_s =$ (13$\pm$1)\% for the signal photons and $D_p =$ (9.4$\pm$0.5)\% for the probe photons. We did not include the PMFs' collection efficiencies when deriving a generation rate from a detection rate of the biphoton source. The inclusion may overestimate the biphoton generation rate because the heralded and heralding photons that are not collected in the optical fiber may not be in pairs. In addition, the biphotons that were collected in the PMFs had well-defined spatial modes. Thus, the generation rate, SB, ESB, quality factor, and pairing probability quoted in this work refer to the biphotons that exit the vapor cell with the spatial modes defined by the PMFs or, equivalently, to those collected inside the PMFs immediately after the vapor cell.

We divided the biphoton detection rate, i.e., the coincidence count per second, by the product of $D_s$ and $D_p$ to obtain the biphoton generation rate $R_b$. Furthermore, we divided the signal (probe) photon detection rate by $D_s$ ($D_p$) to obtain the signal (probe) photon generation rate $R_s$ ($R_p$). The signal (probe) heralding probability $h_s$ ($h_p$) is equal to the ratio of $R_b$ to $R_s$ ($R_b$ to $R_p$), and these heralding probabilities are essential for verifying \eq{relation}. References \onlinecite{OurQST2025} and \onlinecite{OurOpticaQ2025} provide more experimental details.

\section{Results and Discussion}

In this section, we first varied the experimental parameters to understand how the pairing probability changes. We then report the optimal pairing probability achieved in our systematic study and and combine all the data to verify \eq{relation} experimentally. Finally, we conclude that the spectral brightness (SB) indeed has an upper bound described by \eq{relation}.

The experimental parameters that were varied included the pump and coupling field powers and detunings, as these are known to influence the performance—e.g., the generation rate, temporal width, and background photon rate—of the double-$\Lambda$ SFWM biphoton source with the transition diagram shown in \fig{transitions}, as indicated by the biphoton generation theory presented in \SecBiphotonTheory ~\cite{OurAQT2024, BiphotonTheory} and demonstrated in our previous works \cite{OurOPEX2021, OurPRR2022, OurPRA2022, OurOpticaQ2025}. Importantly, all other parameters in the experiments were kept constant. Meanwhile, theoretical simulations were carried out to elucidate the mechanisms underlying the observed variations in the heralding probabilities.

We first studied $h_s$ as a function of the pump power shown by \fig{two}(a), which is proportional to the Rabi frequency squared $\Omega_p^2$, and the pump detuning $\Delta_p$ shown by \fig{two}(b). The red lines are the theoretical predictions based on \EqHs~illustrated in \SecHs, and the overall behavior of $h_s$ of the experimental data agrees with those of the predictions, except for the data point of $\Delta_p$ = 1.4~GHz in \fig{two}(b). We discuss this data point in \SecEquivalence, which can reveal the tendency of decreasing $h_s$ as $\Delta_p$ gets too small. The remaining data points indicate that $h_s$ varied very little with the pump power and detuning; therefore, improving $h_s$ through these parameters is limited. We explain the behavior of $h_s$ in \SecHs.

\FigTwo

We further investigated the probe's heralding probability $h_p$ as a function of the pump power and $\Delta_p$ shown by \figs{two}(a) and \ref{fig:two}(b), respectively. The experimental data (i.e., the blue circles and squares) reveal that $h_p$ increases with increasing pump transition rate (proportional to $\Omega_p^2/\Delta_p^2$), i.e., increases and decreases with increasing pump power and $\Delta_p$, respectively. This behavior is consistent with the theoretical predictions (blue lines in the figures) and is attributed to the ratio of the fluorescence photon to biphoton generation rates. We explain the behavior of $h_p$ in \SecHp.

The effect of tuning the pump power on $h_s$ and $h_p$ in \fig{two}(a) is equivalent to that of tuning $\Delta_p$ in \fig{two}(b). As demonstrated by \FigsSixAandB, this equivalence originates from the roles of the pump power ($\propto \Omega_p^2$) and $\Delta_p$ in the two-photon correlation function, i.e., \EqBiphoton, which determines $R_b$. At a large $\Delta_p$, it can be factored out of the integrals in \EqBiphotonEqFWM. Consequently, $R_b$ is linearly proportional to $\Omega_p^2/\Delta_p^2$ and the proportionality depends on neither $\Omega_p$ nor $\Delta_p$, which facilitates the optimization of the biphoton source performance.

We next studied the signal's heralding probability $h_s$ as a function of the coupling power shown by \fig{three}(a), which is proportional to the Rabi frequency squared $\Omega_c^2$, and a function of the coupling detuning $\Delta_c$ shown by \ref{fig:three}(b). The overall trends of the experimental data (red circles and squares) agree with those of the theoretical predictions (red lines). The data revealed that, as compared with $h_p$, which will be discussed in the next paragraph, $h_s$ slightly varied with the coupling power, whereas $h_s$ exhibited a clear optimum as a function of $\Delta_c$. Such behavior in the data indicates that favorable conditions for $h_s$ can be readily identified, as it is primarily sensitive to $\Delta_c$. A complete explanation of the behavior of $h_s$ is provided in \SecHs.

We further investigated the probe heralding probability $h_p$ as a function of the coupling power ($\propto \Omega_c^2$) and $\Delta_c$ shown in \figs{three}(a) and \ref{fig:three}(b). The overall trends of the experimental data (blue circles and squares) are consistent with those of the theoretical predictions (blue lines). The data revealed that $h_p$ significantly decreased (increased) with the coupling power ($\Delta_c$), and we attribute this behavior to the noise photons in the probe detection channel. A complete explanation of the behavior of $h_p$ is provided in \SecHp.

\FigThree

After knowing how the powers and detunings of the pump and coupling fields affect the heralding probabilities $h_s$ and $h_p$, we aimed to determine the maximum of the pairing probability, i.e., the product of $h_s$ and $h_p$. Seeking the maximum of the pairing probability allows us to get closer to the ideal condition. The four-dimensional search was conducted in the volume formed by the $x$-axis ranges in \figs{two} and \ref{fig:three}. The step sizes for the pump and coupling detunings were 0.5 and 0.1~GHz, and the step sizes for the pump and coupling powers were 0.5 and 1~mW. For a given set of $\Delta_p$, $\Delta_c$, and coupling power, the pump power was typically increased to the maximum value allowed by the single photon criterion that the peak cross-correlation $>$ 7.5 \cite{OurOPEX2024}, i.e., the conditional auto-correlation $<$ 0.5 \cite{Criterion}. This procedure was adopted because, as shown in \fig{two}(a), a higher pump power significantly enhances the $h_p$.

In \fig{four}, we present the biphoton wave packet or two-photon correlation function, i.e., the coincidence count as a function of the delay time of the probe photon upon the heralding by the signal photon, at the highest pairing probability. With the same experimental condition as that of \fig{four}, several measurements on different days revealed that the biphoton source had a generation rate of (8.4$\pm$0.3)$\times$$10^5$ pairs/s and a temporal FWHM of 123$\pm$3~ns, corresponding to an ESB of (1.03$\pm$0.03)$\times$$10^5$~(pairs/s)$\cdot$$\mu$s, and an SBR of 7.1$\pm$0.1. Furthermore, the pairing probability was 0.89$\pm$0.02, with corresponding $h_s$ and $h_p$ of 0.89$\pm$0.02 and 1.00$\pm$0.03. Based on \eq{relation} and with the consideration of uncertainties, the quality factor of 0.73$\pm$0.02, pairing probability of 0.89$\pm$0.02, and $C$ of 0.809$\pm$0.002 agreed well. One point worth mentioning is that the value of $C$ was determined, for example, from the green curve in \fig{four}. The parameter $C$ characterizes the biphoton waveform, and a thorough explanation is provided in \SecQF.

\FigFour

The frequency spectrum of $G^{(2)}(\tau)$ (green line in the inset of \fig{four}) shows an asymmetry in its peak, making it difficult to determine the linewidth. Therefore, we used a Lorentzian function (blue line) to determine the spectrum's linewidth, which has a FWHM slightly greater than that of the green line. Hence, the biphoton generation rate divided by the linewidth indicates an SB of (8.5$\pm$0.3)$\times$$10^5$ pairs/s/MHz. The quality factor and SB here reach 90\% of the fundamental limit. These values are also the highest recorded among all types of HSP sources, as shown by Table~\ref{table:one}.

For an application requiring high fidelity, an SBR of 7.1 can be insufficient. Nevertheless, one can decrease the biphoton temporal width to improve the SBR significantly at the expense of a significant decrease in spectral brightness (SB), even in the case of a noise-free biphoton source. In the hot-atom biphoton source, we can shorten the biphoton temporal width by increasing the power of the coupling laser field \cite{OurPRA2022}. However, an increase in the coupling power slightly decreased the biphoton generation rate and significantly increased the noise photon rate. To demonstrate an approach to the fundamental limit, we used a moderate coupling power resulting in an SBR of 7.1.

\FigFive

We next examined the relationship between the quality factor $Q$ and the pairing probability $P$ as shown in \eq{relation}. The biphoton generation rate $R_b$ the biphoton wave packet SBR and temporal FWHM $\Delta\tau$ and the heralding probabilities $h_s$ and $h_p$ were measured. The product of $R_b$ and $\Delta\tau$ gave the measured ESB, and that of ESB and SBR gave $Q$. Additionally, the product of $h_s$ and $h_s$ gave $P$. The experimental data reveal that the quality factor is linearly proportional to the pairing probability shown in \fig{five}(a). The black straight line with a zero intercept represents the best fit of the data. Due to differences in the biphoton temporal profiles across various experimental conditions, some data points deviate significantly from the black line.

To account for these differences, we calculated the proportionality $C$ in \eq{relation} for each biphoton temporal profile. Then, we plotted the normalized quality factor, i.e., $Q/C$, as a function of $P$ in \fig{five}(b). The straight line with a slope of 1 represents the theoretical prediction that $Q = C$$\times$$P$. All the experimental data points align well with the theoretical prediction, verifying the relationship between $Q$ and $P$. Note that the detection efficiencies $D_s$ and $D_p$ do not influence the verification of the relationship $Q = C$$\times$$P$ presented in \fig{five}(b) (see the last paragraph in \SecQF~ for details). The experimental confirmation of this relationship demonstrates that SB indeed has an upper bound, which is determined by how closely the source approaches the noise-free condition and the corresponding SBR.

Although we carried out the study yielding the results in \fig{five} with a hot-atom double-$\Lambda$ SFWM biphoton source, \eq{relation} also holds for biphoton sources on different platforms, whether via SPDC or SFWM. Only a few references provided the $h_s$ and $h_p$ data of their biphoton sources, and the $Q$ and $P$ values for these references satisfied \eq{relation} \cite{\CiteQualityFactor}. Compared with other values listed in Table~\ref{table:one}, our data in Fig.~\ref{fig:four} exhibit the highest $P$ value, indicating that our hot-atom biphoton source is the closest to an ideal near–noise-free source. This performance is achieved by exploiting coupling detuning to significantly enhance $h_s$, together with systematic optimization of the pump and coupling fields to increase biphoton generation while suppressing fluorescence in the probe channel.

\section{Conclusion}

The spectral brightness (SB), a key metric for heralded single-photon (HSP) or biphoton sources, has long been considered a quantity to maximize, yet whether it has a fundamental upper limit remains uncertain. To resolve this upper limit, we introduced the universal relationship between the quality factor $Q$ and the pairing probability $P$, outlined in \eq{relation}. We systematically studied how various experimental parameters affect $P$ with hot atomic vapor and provided theoretical interpretations of the observed behaviors. The experimental data presented in \fig{five}(b) precisely verify the formula of $Q = C$$\times$$P$. The physics underlying the formula is depicted as follows: For an ideal noise-free biphoton source, i.e., $P = 1$, the SBR degrades as the probability of overlap between biphoton pairs increases, which occurs when the $R_b$, $\Delta\tau$, or their product, i.e., the effective spectral brightness (ESB), increases. Therefore, a higher SBR is achieved at the cost of a lower ESB or SB. With a nonideal biphoton source, i.e., $P < 1$, in which noise can be uncorrelated signal photons, uncorrelated probe photons, fluorescence photons, photons due to laser leakages, etc., the degradation is more severe. 

After optimizing various experimental parameters, we achieved a $P$ value of 0.89$\pm$0.02; i.e., approximately $94\%$ of the photons in the two detection channels were, on average, paired biphotons. This $P$ value resulted in a $Q$ value of 0.73$\pm$0.02, corresponding to an ESB of (1.03$\pm$0.03)$\times$$10^5$ (pairs/s)$\cdot$$\mu$s or an SB of (8.5$\pm$0.3)$\times$$10^5$ pairs/s/MHz at an SBR of 7.1$\pm$0.1, referring to the biphotons collected inside the single-mode PMFs immediately after the atomic vapor cell. The quality factor and its SB are the highest among all biphoton sources with different generation processes and various media that satisfy the single-photon criterion (Table~\ref{table:one}). This result marks a significant milestone in the development of HSP or biphoton sources. In future work, we aim to increase $P$ toward unity further and achieve the maximum SB for arbitrary SBRs, enabling optimal biphoton performance independent of SBR.

\section*{Data availability statement}
All data that support the findings of this study are included within the paper. 

\section*{Acknowledgments}
This work was supported by the National Science and Technology Council, Taiwan, under grants 112-2112-M-007-020-MY3 and 114-2119-M-007-012, and by the Taiwan Centers of Excellence program of the Ministry of Education, Taiwan. I.A.Y. thanks Dr. Thorsten Peters for the fruitful discussion.

\section*{Data availability}
Data underlying the results presented in this paper are not publicly available at this time but may be obtained from the authors upon reasonable request.

\section*{Funding}
National Science and Technology Council, Taiwan (112-2112-M-007-020-MY3, 114-2119-M-007-012).

\section*{Conflict of interest}
The authors declare no conflicts of interest.

\appendix

\section{Relation between the quality factor and pairing probability}
\label{app:QF}

Equation~(1) in the main text shows the relation between the quality factor $Q$ and pairing probability $P$, i.e., $Q = C$$\times$$P$ where $C$ is a dimensionless proportionality depending only on the biphoton wave packet's shape. \RFs{Table16, Formula1, Formula2, Formula3} showed formulas that relate a coincidence-to-accidental ratio (CAR) to the corresponding biphoton generation rate, where CAR is the number ratio of coincidence counts to accidental counts within the time window for recording the heralded photon's arrival. Since those formulas contain no information about the biphoton temporal width, they do not reveal the underlying physics that the fundamental limit of $Q$ is due to the overlap between biphoton wave packets. Considering the biphoton temporal width, we derive $Q =  C$$\times$$P$ here.

We first illustrate a simple case of the square-pulse biphoton wave packet with a temporal width of $\Delta\tau$. The SBR $r_{SB}$ is given by the ratio of the correlated $\langle N_c \rangle$ and uncorrelated $\langle N_{unc} \rangle$ photon numbers per time bin per heralding event, i.e.,
\begin{equation}
\label{eq:SBR}
	r_{SB} = \frac{\langle N_c(\tau_0)  \rangle}{\langle N_{unc} \rangle}
		= \frac{h_s \times (\Delta t_{\rm bin}/\Delta\tau) + (1-h_s) \times 0}
		{R_p \Delta t_{\rm bin}},
\end{equation}
where $\tau_0$ represents the peak position of the biphoton wave packet, and $\Delta\tau$ denotes the biphoton temporal FWHM, $h_s$ is the signal photon's heralding probability, $R_p$ is the probe photon generation rate, and $\Delta t_{\rm bin}$ is the time bin width. In the numerator of the above equation, $h_s \times (\Delta t_{\rm bin}/\Delta\tau)$ means that each photon in the heralding channel has a probability of $h_s$ to pair with its counterpart or heralded photon, and the counterpart photon has a probability of $\Delta t_{\rm bin}/\Delta\tau$ to appear in the time bin centering around $\tau_0$ due to the square pulse. Naturally, each photon in the heralding channel has a probability of $1-h_s$ without any counterpart or does not contribute to $\langle N_c(\tau_0)  \rangle$. The above equation now becomes
\begin{equation}
	r_{SB} = \frac{h_s /\Delta\tau}{R_b / h_p} = \frac{h_s h_p}{R_b \Delta\tau}
\end{equation}
or, equivalently,
\begin{equation}	
	Q \equiv r_{SB} (R_b\Delta\tau)  = h_s h_p \equiv P.
\end{equation}
Thus, $Q = P$ for the square-pulse biphoton wave packet.

Following the above derivation, we now consider the biphoton wave packet with an arbitrary pulse shape. The term of $\Delta t_{\rm bin}/\Delta\tau$ in the numerator of \eq{SBR} is revised to $C\times(\Delta t_{\rm bin}/\Delta\tau)$, where
\begin{equation}
\label{eq:C_definition}
	C \equiv \left[ \frac{f(\tau_0)}{\int^{\infty}_{-\infty} f(\tau) d\tau} \right] \Delta\tau
\end{equation}
and $f(\tau)$ is the biphoton wave packet's temporal profile. The expression in the brackets represents the peak value of a normalized biphoton wave packet, where the normalization requires the probability of the entire wave packet being 1. The variation in the $C$ value is solely due to different temporal shapes, which yield different peak values after normalization. Consequently, we have obtained
\begin{equation}
\label{eq:QCPinAppA}
	Q = C \times P.
\end{equation}
In the derivation, we utilize the condition that the time bin width is much shorter than the biphoton temporal width, i.e., $\Delta t_{\rm bin} \ll \Delta\tau$, and the approximation of one heralded photon per successful heralding event, i.e., $h_s \times (\Delta t_{\rm bin}/\Delta\tau)$ for square pulses in \eq{SBR} or $h_s \times C (\Delta t_{\rm bin}/\Delta\tau)$ for arbitrary pulses. The above equation is valid for all biphoton sources, regardless of their generation processes or platforms. Table~\ref{table:three} shows the values of $C$ of given temporal shapes or, equivalently, the maximum values of $Q$ at $P = 1$. 

\TableThree

The relation of $Q = C$$\times$$P$ in \eq{QCPinAppA} does not depend on the detection efficiencies $D_s$ and $D_p$ of photons in the heralding and heralded channels. One multiplies $D_s D_p$ to both sides of the equation and obtains
\begin{equation}
	(R_b D_s D_p) \, \Delta\tau \, r_{SB} = C \times (h_s D_p) (h_p D_s)
\end{equation}
or, equivalently,
\begin{equation}
	R_d \, \Delta\tau \, r_{SB} = C \times e_s e_p,
\end{equation}
where $R_d$ is the biphoton detection rate or coincidence count rate, and $e_s$ ($e_p$) is the heralding efficiency by using the detected signal (probe) photons to herald detecting coincidence counts. The biphoton source's $R_d$, $e_s$, and $e_p$ are the directly measured quantities. Both of $\Delta\tau$ and $r_{SB}$ are independent of $D_s$ and $D_p$. Therefore, the detection efficiencies do not influence the verification of $Q =  C$$\times$$P$ presented in \FigFiveB.

\section{Theoretical predicitons of the two-photon correlation function and biphoton generation rate}
\label{app:BiphotonTheory}

We use the biphoton generation rate in the theoretical predictions of the signal's and probe's heralding probabilities. In this appendix, we illustrate how to calculate the biphoton generation rate from the two-photon correlation function $G^{(2)}(\tau)$, i.e., the biphoton wave packet \cite{BiphotonTheory, OurAQT2024}. \RF{OurOpticaQ2025} provides the study on $G^{(2)}(\tau)$ based on the double-$\Lambda$ SFWM biphoton source using a Doppler-broadened atomic medium. For the completeness of this article, but without illustration in depth, we list the formula of $G^{(2)}(\tau)$ below:
\begin{eqnarray}
\label{eq:biphoton}
	G^{(2)}(\tau) = 
		\left| 
		\int_{-\infty}^{\infty} d\delta \frac{e^{-i\delta\tau}}{2\pi}
		\bar{\kappa}(\delta)\,
		{\rm sinc} [\bar{\rho}_c(\delta) + \bar{\rho}_m(\delta)]
		e^{i[\bar{\rho_c}(\delta) + \bar{\rho}_m(\delta)]}
		B(\delta)
		\right|^2,
\end{eqnarray}
where $\delta$ denotes the two-photon detuning, $\bar{\kappa}(\delta)$ is proportional to the cross-susceptibility between the signal and probe photons, $\bar{\rho}_c(\delta)$ and $\bar{\rho}_m(\delta)$ are proportional to the probe photon self-susceptibilities induced by the atoms exhibiting the electromagnetically induced transparency (EIT) effect and those without the EIT effect or the impurity atoms, respectively, and $B(\delta)$ represents the combined frequency spectrum of the probe and signal etalons. The four functions of $\delta$ are defined as follows:
\begin{eqnarray}
\label{eq:FWM}
	\hspace*{-20mm} \bar{\kappa}(\delta) &&=
		\frac{(1-b)\alpha}{4}\int_{-\infty}^{\infty} d\omega_D
		\left[  
		\frac{e^{-\omega_D^2/\Gamma_D^2}}{\sqrt{\pi}\Gamma_D}
		\frac{\Omega_p}{\Delta_p + \omega_D + i\Gamma/2}
		\right.
		\nonumber \\
	\hspace*{-20mm} &&\times
		\left.
		\frac{\Omega_c\Gamma}{\Omega_c^2-4(\delta+i\gamma)
		(\delta+\Delta_c+\omega_D+i\Gamma/2)}
		\right],
		\\
\label{eq:EITc}
	\hspace*{-20mm} \bar{\rho}_c(\delta) &&=
		\frac{(1-b)\alpha}{2} \int_{-\infty}^{\infty} d\omega_D
		\left[
		\frac{e^{-\omega_D^2/\Gamma_D^2}}{\sqrt{\pi}\Gamma_D}
		\frac{(\delta+i\gamma)\Gamma}
		{\Omega_c^2-4(\delta+i\gamma)(\delta+\Delta_c+\omega_D+i\Gamma/2)}
		\right],
		\\
\label{eq:EITm}
	\hspace*{-20mm} \bar{\rho}_m(\delta) &&=
		\frac{b\alpha}{2} \int_{-\infty}^{\infty} d\omega_D
		\left[
		\frac{e^{-\omega_D^2/\Gamma_D^2}}{\sqrt{\pi}\Gamma_D}
		\frac{\Gamma}
		{4(\delta+\Delta_c+\omega_D+i\Gamma/2)}
		\right],
		\\
\label{eq:etalon}
	\hspace*{-20mm} B(\delta) &&=
		\left( \frac{1}{1+2i\delta/\Gamma_e} \right)^4,
\end{eqnarray} 
where $\omega_D$ represents the Doppler shift due to atomic velocity, $\alpha$ is the optical depth (OD) of the medium, $b$ is the fraction of the impurity atoms, $\Gamma$ ($\approx 2\pi\times$6~MHz in this work) is the spontaneous decay rate of the excited states, $\Omega_p$ and $\Omega_c$ denote the Rabi frequencies of the pump and coupling fields, $\Delta_p$ and $\Delta_c$ are the pump's and coupling's one-photon detuning, $\gamma$ represents the decoherence rate in the experimental system, and $\Gamma_e$ is the spectral FWHM of each etalon. 

The biphoton generation rate $R_b$ is proportional to the integral of the biphoton wave packet over the delay time given by
\begin{equation}
\label{eq:BGR}
	R_b \propto \int_{-\infty}^{\infty} d\tau G^{(2)}(\tau)~~{\rm or}~~
		R_b = A \int_{-\infty}^{\infty} d\tau G^{(2)}(\tau),
\end{equation} 
where $A$ is a dimensionless proportionality accounting for the ratio of the measured value of $R_b$ to the theoretical value of $\int_{-\infty}^{\infty} d\tau G^{(2)}(\tau)$. The day-to-day variations of the detection efficiencies affected $A$. We kept $A$ constant to match the predictions of $h_s$ and $h_p$ to the experimental data in each of \FigsTwoThreeAll. The measured values of $R_b$ revealed that $A$ varied within (1.3$\pm$0.1)$\times$$10^6$ between the figures.

In the calculations of $G^{(2)}(\tau)$ and $R_b$, we set $\Gamma_D$ = 54$\Gamma$ based on the atomic vapor cell's temperature, $\alpha = 500$ determined from the absorption spectrum of a weak laser field, and $\Gamma_e$ (= 2$\pi$$\times$53.5~MHz or 8.9$\Gamma$) determined by measuring the signal and probe etalons' transmission spectra. Furthermore, $\Omega_c$, $b$, and $\gamma$ significantly influenced $R_b$ and the biphoton temporal width $\Delta\tau$, and we utilized a series of experimental data of $R_b$ and $\Delta\tau$ as functions of the coupling detuning or power to determine their values \cite{OurOpticaQ2025}. After obtaining the ratio of $\Omega_c^2$ to the coupling power and the value of $b$ that best matched the predictions with a constant $\gamma$ to the data, we slightly adjusted $\gamma$ to make each predicted temporal profile better agree with its corresponding experimental one \cite{OurPRA2022}. Since $\Omega_p$ does not affect the temporal profile and only changes the value of $A$ in \eq{BGR} \cite{OurPRR2022}, its absolute magnitude is irrelevant in this study. We estimated the value of $\Omega_p$ with the power, beam size, and transition strength of the pump field \cite{DataSheet, OurOPEX2021}.

\section{The signal's heralding probability}
\label{app:Hs}

We start with the illustration of the formula for the signal's heralding probability $h_s$. The photon generation rate in the signal detection channel $R_s$ is proportional to the pump-signal two-photon transition rate. In addition to the biphotons' signal photons, the photons resulting in $R_s$ also consist of the signal photons that have lost their counterparts (i.e., the uncorrelated signal photons), fluorescence photons induced by the pump field, and pump laser leakage. In this work, the latter two made negligible contributions to $R_s$ with a large one-photon detuning. The transition rate at the two-photon resonance is approximately proportional to $(\Omega_p/2\Delta_p)^2$, where $\Omega_p$ and $\Delta_p$ represent the pump Rabi frequency and detuning. Thus, we obtain
\begin{equation}
\label{eq:hs}
	h_s \equiv \frac{R_b}{R_s} = \frac{R_b}{B[\Omega_p^2/(4\Delta_p^2)]},
\end{equation}
where $B$ is a proportionality and the biphoton generation rate $R_b$ is given by \eq{BGR}. $B$ has no dependence on $\Omega_p$, $\Delta_p$, or the coupling power or detuning.

Similar to the proportionality $A$ in $R_b$, we kept the value of $B$ constant in each of \FigsTwoThreeAll. Note that $A$ and $B$ are not two independently adjustable parameters in \eq{hs}, but $A/B$ is the only one to match the overall magnitude of predictions to that of data. The best matches revealed that the values of $A/B$ varied within (1.4$\pm$0.1)$\times$$10^{-4}$ s$^{-1}$ between the figures.

We explain the behavior of $h_s$ versus the pump power and detuning, as shown by \FigsTwoAandB, in this and the next paragraphs. $R_b$ is linearly proportional to the pump transition rate, i.e., $\approx \Omega_p^2/\Delta_p^2$ at large $\Delta_p$, as expected from the biphoton generation theory. $R_s$ is also linearly proportional to the pump transition rate due to the spontaneous Raman transition induced by the pump field and signal photon. Hence, $h_s$ (= $R_b / Rs$) varied very little with the pump power and detuning.

The slight variations of $h_s$ versus the pump power and detuning were caused by the variation of the ground-state decoherence rate $\gamma$. Since $\gamma$ also affected a biphoton wave packet's temporal width, we utilized \eq{biphoton} of \SecBiphotonTheory\ and compared the theoretical and experimental temporal profiles to determine the value of $\gamma$ at each pump power and detuning. The red lines in \FigsTwoAandB\ are the theoretical prediction of $h_s$ calculated with the pre-determined value of $\gamma$ and \eq{hs}. The theoretical predictions will be two horizontal lines if we set $\gamma$ equal for all pump powers and detunings. Therefore, as long as the pump power or detuning is the only variable in the hot-atom double-$\Lambda$ SFWM biphoton source, it does not affect the signal's heralding probability.

Regarding the behavior of $h_s$ against the coupling power in \FigThreeA, the FWM efficiency, i.e., $|\bar{\kappa}|^2$ [see \eq{biphoton}], and the variation of the ground-state decoherence rate $\gamma$ are the two main effects. The fraction involving $\Omega_c$ in $|\bar{\kappa}|^2$ [see \eq{FWM}] illustrates that the biphoton generation rate $R_b$ increases with the numerator's $\Omega_c^2$, when the term of $\Omega_c^2$ was smaller than that of $8\gamma\Delta_c^2/\Gamma$ in the denominator. The experimental condition indeed satisfied $\Omega_c^2 <$ $8\gamma\Delta_c^2/\Gamma$. Furthermore, the pre-determined value of $\gamma$ increased with the coupling power, and a more significant $\gamma$ resulted in a smaller $R_b$. The combined effect of $|\bar{\kappa}|^2$ and $\gamma$ results in an optimum $R_b$ against coupling power. In \FigThreeA, the pump power and detuning were constant, and the signal photon generation rate $R_s$ did not change. Hence, $h_s$ ($= R_b/R_s$) slightly varied with the coupling power and had an optimum value against it.

Regarding the behavior of $h_s$ against $\Delta_c$ in \FigThreeB, the non-EIT atoms and $|\bar{\kappa}|^2$ are the two main effects. We did not show the data point for $\Delta_c/2\pi <$ 0.5~GHz in the figure, because as $\Delta_c$ approached zero, $h_s$ dropped dramatically due to the absorption of non-EIT atoms (see the study in Ref.~\cite{OurOpticaQ2025}). Note that in this work, the absorption coefficient is reduced to 0.1 at $\Delta_c/2\pi =$ 630~MHz. As $\Delta_c \geq$ 0.5~GHz, it decreased the FWM efficiency as indicated by the term of $\gamma^2\Delta_c^2$ in the denominator of $|\bar{\kappa}|^2$. Note that the pre-determined $\gamma$ values were about the same against $\Delta_c$. The theoretical predictions calculated from \eq{biphoton} consider all of the effects on $R_b$, including the above-mentioned non-EIT atoms and FWM efficiency, due to the coupling detuning. Since the pump power and detuning were constant, $R_s$ remained the same. Thus, for $\Delta_c/2\pi \geq$ 0.5~GHz, $h_s = (R_b/R_s)$ changed slightly compared with $h_p$, and the value of $h_s$ first increased and then decreased with $\Delta_c$.

\section{The probe's heralding probability}
\label{app:Hp}

We start with the illustration of the formula for the probe's heralding probability $h_p$. The photon generation rate in the probe detection channel $R_p$ consists of two main parts. The first part is the biphoton's or correlated probe photon rate, equal to $R_b$. The second part is the rate of noise photons $R_{pn}$ appearing without the presence of the pump field. Although we applied the hollow-cored pumping field to optically pump out of the ground state $|2\rangle$ driven by the coupling field, a few atoms entering the SFWM interaction region still had the population in $|2\rangle$. These atoms and the coupling field contributed to the noise photons. Thus, $R_p = R_b + R_{pn}$ and we obtain
\begin{equation}
\label{eq:hp}
	h_p \equiv \frac{R_b}{R_p} = \frac{R_b}{R_b + R_{pn}},
\end{equation}
where $R_b$ is given by \eq{BGR}.

The coupling power ($\propto$ $\Omega_c^2$) and detuning ($\Delta_c$) and other experimental conditions can influence $R_{pn}$. Since the proportionality $A$ in $R_b$ is already an adjustable parameter to match the overall magnitude of theoretical predictions to that of experimental data, we intend not to introduce another independently adjustable parameter in the formula of $h_p$. Thus, $R_{pn}$ in \eq{hp} is the measured value in the corresponding experimental condition.

We explain the behavior of $h_p$ versus the pump power and detuning, as shown by \FigsTwoAandB, in this and the next paragraphs. The behavior was caused by noise photons in the probe detection channel, which existed even in the absence of the pump field. While the biphoton generation rate $R_b$ increases with the pump transition rate, the noise photon rate is approximately independent of it. A higher pump power ($\Delta_p$) makes the noise photons more (less) significant relative to the total photons in the probe detection channel. Consequently, $h_p$ significantly increases (decreases) with the pump power ($\Delta_p$).

To verify the above explanation, we measured the noise photon rate in the probe detection channel without the presence of the pump field. The theoretical prediction of $h_p$ is based on the measured noise photon rate. The blue lines in \FigsTwoAandB\ are the theoretical predictions of $h_p$ based on \eq{hp}. The consistency between the predictions and the data is satisfactory, clarifying that the ratio of the noise photon rate in the probe detection channel to the rate of biphotons' probe photons is the decisive factor for the probe's heralding probability in the hot-atom double-$\Lambda$ SFWM biphoton source.

In \FigsThreeAandB, the behavior of $h_p$ significantly decreasing or increasing with the coupling power or $\Delta_c$, respectively, can again be explained by the noise photons in the probe detection channel. A larger coupling power ($\Delta_c$) induced more (fewer) fluorescence photons, and the noise photon rate approximately linearly increased (decreased) with the coupling power ($\Delta_c$). On the other hand, $R_b$ varied slightly with the coupling power ($\Delta_c$) as demonstrated by the insignificant changes of $h_s$ against the coupling power ($\Delta_c$) in the exact figure illustrated earlier. Consequently, the coupling power ($\Delta_c$) in the region of the figure reduced (enhanced) $h_p$ significantly.

\section{Equivalence between the effects on the heralding probabilities induced by the pump power and detuning}
\label{app:Equivalence}

\FigSix

We present the experimental evidence in this appendix, clarifying that the effect on the heralding probabilities $h_s$ and $h_p$ induced by the pump power $P_p$ ($\propto \Omega_p^2$) is equivalent to that by the pump detuning $\Delta_p$. \FIGs{six}(a) and \ref{fig:six}(b) show $h_s$ and $h_p$ as functions of $P_p/\Delta_p^2$, respectively. One can either increase $P_p$ or decrease $\Delta_p^2$ to achieve the same values of $h_s$ and $h_p$. The effects due to $P_p$ and $\Delta_p^2$ are equivalent.

The biphoton generation theory of the double-$\Lambda$ SFWM source in \SecBiphotonTheory\ also supports the equivalence. As long as $\Delta_p$ is significantly greater than the natural linewidth $\Gamma/2\pi$ ($\approx$ 6~MHz) and also the Doppler width $\Gamma_D/2\pi$ ($\approx$ 320~MHz), one can move the term containing $\Omega_p$ and $\Delta_p$ in \eq{FWM} out of the integral or ensemble average of a Doppler-broadened medium, i.e.,
\begin{equation}
	\bar{\kappa}(\delta) \approx
		\frac{\Omega_p}{\Delta_p} \, \bar{\kappa}^{\prime}(\delta),
\end{equation}
where $\bar{\kappa}^{\prime}(\delta)$ represents the remaining part of $\bar{\kappa}(\delta)$ that does not consist of $\Omega_p$ and $\Delta_p$. Then, the two-photon correlation function in \eq{biphoton} and the biphoton generation rate in \eq{BGR} become
\begin{equation}
	G^{(2)}(\tau) \approx 
		\frac{\Omega_p^2}{\Delta_p^2} G^{\prime(2)}(\tau)
\end{equation}
and
\begin{equation}
	R_b \approx \frac{\Omega_p^2}{\Delta_p^2}~
		A \int_{-\infty}^{\infty} d\tau G^{\prime(2)}(\tau),
\end{equation} 
where $G^{\prime(2)}(\tau)$ denotes the remaining part of $G^{(2)}(\tau)$ that does not consist of $\Omega_p$ and $\Delta_p$. Thus, $R_b$ is linearly proportional to $\Omega_p^2/\Delta_p^2$. Finally, $h_s$ in \eq{hs} and $h_p$ in \eq{hp} are functions of a single independent variable of $\Omega_p^2/\Delta_p^2$ but not two independent ones of $\Omega_p$ and $\Delta_p$. The nearly complete overlap between the solid and dashed lines in \figs{six}(a) and \ref{fig:six}(b) verifies the above arguments. Hence, the effects on the heralding probabilities from increasing the pump power and from decreasing the pump detuning are equivalent.

The rightmost squared data point in \fig{six}(a) significantly deviates from the theoretical prediction. This data point's $\Delta_p/2\pi$ is 1.4~GHz. We attributed the deviation to fluorescence photons induced by the pump transition to the excited state $|4\rangle$. The fluorescence rate cannot be formulated by $\Omega_p^2/(4\Delta_p^2)$, i.e., the pump transition rate at the zero atomic velocity, when $\Delta_p$ is not significantly larger than the Doppler width. One needs to do the velocity average of the pump transition rate to obtain the fluorescence rate, and the average value is larger than the value of $B \Omega_p^2/(4\Delta_p^2)$. Furthermore, a smaller $\Delta_p$ resulted in a higher transmission of fluorescence photons through the etalons, leading to more noise photons in the signal detection channel. As $\Delta_p$ got small, the fluorescence-induced noise photon rate became nonnegligible compared to the signal photon rate, reducing the signal's heralding probability.

\secRefs

\end{document}